\documentclass[twocolumn]{aastex631}
\usepackage{natbib}

\begin{document}

\title{A First Order Filter for the Detection of Potentially Habitable Exoplanets}

\author{Raka Dabhade}
\affiliation{Department of Physics \\
Fergusson College (Autonomous) \\
Shivajinagar, Pune, India}

\author{Jebraan Mudholkar}
\affiliation{Department of Physics \\
Fergusson College (Autonomous) \\
Shivajinagar, Pune, India}

\author{Siddhesh Durgude}
\affiliation{Department of Physics \\
Fergusson College (Autonomous) \\
Shivajinagar, Pune, India}

\author{Arpit Kottur}
\affiliation{Department of Physics \\
Fergusson College (Autonomous) \\
Shivajinagar, Pune, India}

%\collaboration{20}{(AAS Journals Data Editors)}

%\author{F.X Timmes}
%\affiliation{Arizona State University}
%\affiliation{AAS Journals Associate Editor-in-Chief}

%\altaffiliation{AASTeX v6+ programmer}

%\affiliation{AAS Director of Publishing}
%\affiliation{American Astronomical Society \\
%1667 K Street NW, Suite 800 \\



\begin{abstract}
The search for potentially habitable exoplanets is a primary objective in modern astrophysics, yet the vast number of candidates discovered by missions like Kepler and TESS presents a significant challenge for detailed follow-up characterization. An efficient and reliable method for prioritizing the most promising targets is therefore essential. In this paper, we propose a novel first-order filter for identifying potentially habitable worlds based on a simple geometric ratio: the orbital semi-major axis to the stellar diameter ($d/D_s$). Using data from the NASA Exoplanet Archive, we demonstrate that the ideal value for this ratio is not constant, but is dependent on the host star's spectral class. We establish a tiered framework of ideal ratios, beginning with $\approx 108$ for G-type stars (anchored by the Earth-Sun system), and decreasing by a factor of two for K-type ($\approx 54$) and M-type ($\approx 27$) stars, respectively. Our analysis reveals a strong correlation, showing that exoplanets whose $d/D_s$ ratios are close to these empirically derived values consistently exhibit high Earth Similarity Index (ESI) scores. We propose that these tiered ratios represent "Habitability Main Sequences," analogous to the Hertzsprung-Russell diagram for stars, providing a valuable and straightforward tool for the astronomical community to rapidly screen large datasets and efficiently shortlist high-priority candidates for further investigation with next-generation observatories.
\end{abstract}




\section{Introduction}
\label{sec:intro}

\subsection{The Expanding Catalog of Exoplanets}
The last three decades have transformed the study of planetary science from a discipline focused on our solar system to a galactic-scale endeavor. The discovery of the first exoplanet orbiting a sun-like star, 51 Pegasi b \citep{Mayor1995}, heralded a new era in astronomy. This initial discovery was followed by a steady stream of detections from ground-based surveys, but the field was revolutionized by dedicated space-based transit photometry missions. NASA's Kepler Space Telescope, in particular, provided an unprecedented wealth of data, revealing that planets are a common feature of the Milky Way galaxy and identifying thousands of planet candidates \citep{Borucki2010}. The Kepler mission demonstrated that small, potentially rocky planets are numerous and provided the first statistical census of the planet population in our local stellar neighborhood.

More recently, the Transiting Exoplanet Survey Satellite (TESS) has continued this work with an all-sky survey, focusing on bright, nearby stars that are ideal targets for follow-up characterization \citep{Ricker2015}. The combined legacy of these missions has resulted in a catalog of over 5,000 confirmed exoplanets, exhibiting a staggering diversity in mass, size, and orbital architecture. This large and growing dataset presents both an opportunity and a challenge: how to efficiently sift through thousands of candidates to identify those most likely to be habitable.

\subsection{Defining and Identifying Habitability}
The search for life beyond Earth is fundamentally guided by the concept of the circumstellar habitable zone (HZ), colloquially known as the "Goldilocks zone." The classical HZ is defined as the region around a star where a terrestrial planet with a suitable atmosphere could maintain liquid water on its surface \citep{Kasting1993}. The location and width of the HZ are primarily dependent on the luminosity and spectral type of the host star; hotter, more luminous stars have wider HZs located farther out, while cooler, dimmer stars have narrower HZs situated much closer in.

While position within the HZ is a necessary first-order condition for habitability, it is not sufficient. A planet's potential to host life is a complex function of many additional factors, including its mass, radius, atmospheric composition, and geological activity. To quantify a planet's potential based on its physical properties, metrics such as the Earth Similarity Index (ESI) have been developed \citep{SchulzeMakuch2011}. The ESI provides a scale from 0 to 1, comparing a planet's bulk properties (such as radius and incident stellar flux) to those of Earth. Planets with an ESI greater than 0.8 are generally considered "Earth-like" and represent prime targets in the search for habitable worlds.

\subsection{The Need for a First-Order Filter}
The sheer volume of exoplanet candidates necessitates the development of efficient and robust methods for prioritizing targets for detailed follow-up observations, such as atmospheric characterization with the James Webb Space Telescope (JWST). While complex models can assess habitability, they often require data that are not readily available for most candidates. There is therefore a significant need for a simple, physically motivated "first-order filter" that can be applied to large datasets to rapidly identify a manageable list of high-priority candidates.

In this paper, we propose such a filter. We hypothesize that a simple geometric ratio—the orbital semi-major axis to the stellar diameter ($d/D_s$)—can serve as a powerful initial proxy for habitability, provided it is calibrated for the host star's spectral type. We demonstrate that for G, K, and M-type main-sequence stars, the ideal $d/D_s$ ratio for planets with high ESI values decreases by a consistent factor of approximately two for each successive spectral class. We present this relationship in a diagram analogous to the Hertzsprung-Russell diagram, which reveals distinct "habitability sequences" for different stellar classes. This method provides a novel and intuitive tool for the rapid identification of potentially habitable exoplanets.

\section{Theoretical Framework}
\label{sec:theory}

\subsection{Planetary Equilibrium Temperature}
The surface temperature of a planet is a primary determinant of its habitability. A planet's global average temperature can be approximated by its effective or equilibrium temperature ($T_{\text{eq}}$), which is determined by the balance between the energy it absorbs from its host star and the energy it radiates back into space. Assuming the planet radiates as a blackbody, this balance can be expressed as:

\begin{equation}
    4\pi R_p^2 \sigma T_{\text{eq}}^4 = \frac{L_*}{4\pi d^2} \pi R_p^2 (1 - A)
    \label{eq:energy_balance}
\end{equation}

where $R_p$ is the planet's radius, $\sigma$ is the Stefan-Boltzmann constant, $L_*$ is the stellar luminosity, $d$ is the semi-major axis of the planet's orbit, and $A$ is the planetary Bond albedo (the fraction of incident radiation reflected to space). Simplifying this equation gives the standard expression for equilibrium temperature \citep{Pierrehumbert2010}:

\begin{equation}
    T_{\text{eq}} = \left( \frac{L_*(1 - A)}{16\pi\sigma d^2} \right)^{1/4}
    \label{eq:Teq_L}
\end{equation}

For a planet to be considered habitable in an Earth-like context, its temperature must allow for the existence of liquid water, typically considered to be in the range of 273 K to 373 K.

\subsection{The Role of Stellar Properties}
The equilibrium temperature of a planet is critically dependent on the properties of its host star, primarily its luminosity ($L_*$). The luminosity of a main-sequence star is a function of its effective temperature ($T_*$) and its radius ($R_*$), as described by the Stefan-Boltzmann law \citep{Kippenhahn2012}:

\begin{equation}
    L_* = 4\pi R_*^2 \sigma T_*^4
    \label{eq:stefan_boltzmann}
\end{equation}

Substituting Equation \ref{eq:stefan_boltzmann} into Equation \ref{eq:Teq_L} allows us to express the planet's equilibrium temperature directly in terms of the star's physical properties:

\begin{equation}
    T_{\text{eq}} = T_* \left( \frac{R_*}{2d} \right)^{1/2} (1 - A)^{1/4}
    \label{eq:Teq_star}
\end{equation}

This relationship makes it clear that for a fixed orbital distance $d$, a planet orbiting a cooler, smaller star will be significantly colder than one orbiting a hotter, larger star. This is the fundamental reason why the habitable zone is located at different distances for different spectral types \citep{Kasting1993}.

\subsection{Deriving the d/D\textsubscript{s} Ratio as a Habitability Proxy}
Our proposed filter is based on the ratio of the orbital semi-major axis to the stellar diameter ($d/D_s$), where $D_s = 2R_*$. We can rearrange Equation \ref{eq:Teq_star} to isolate this ratio and understand its physical significance. Squaring both sides and solving for the ratio $d/R_*$ yields:

\begin{equation}
    \frac{d}{R_*} = \frac{1}{2} \left( \frac{T_*}{T_{\text{eq}}} \right)^2 (1 - A)^{1/2}
\end{equation}

By substituting $D_s/2$ for $R_*$, we arrive at an expression for our proposed ratio:

\begin{equation}
    \frac{d}{D_s} = \frac{1}{4} \left( \frac{T_*}{T_{\text{eq}}} \right)^2 (1 - A)^{1/2}
    \label{eq:dDs_ratio}
\end{equation}

Equation \ref{eq:dDs_ratio} provides the theoretical foundation for our method. It demonstrates that if we assume a target equilibrium temperature ($T_{\text{eq}}$) conducive to life (e.g., ~288 K for Earth) and a relatively constant albedo ($A$), the ideal $d/D_s$ ratio for a habitable planet is directly proportional to the square of the stellar effective temperature ($T_*^2$). Therefore, as one moves from hotter G-type stars to cooler K-type and M-type stars, the value of $T_*$ decreases, and consequently, the ideal $d/D_s$ ratio required to maintain a clement temperature must also decrease. This provides a clear physical justification for our empirical finding that the optimal ratio is not constant across spectral classes.

\section{Methodology}
\label{sec:methodology}

\subsection{Data Acquisition and Curation}
The foundation of this study is the comprehensive dataset provided by the NASA Exoplanet Archive, accessed via its online data portal \citep{Akeson2013}. We programmatically queried the Planetary Systems Composite Data table to acquire the most recent and complete catalog of confirmed exoplanets. The initial dataset comprised over 5,000 entries. For the purpose of our analysis, a specific set of physical and orbital parameters was required for each planet-star system: the planet's orbital semi-major axis ($d$), the planetary radius ($R_p$), the host star's radius ($R_*$, from which diameter $D_s = 2R_*$ is derived), and the host star's effective temperature ($T_*$).

A rigorous data curation process was then applied. All entries that lacked definitive values for any of these four key parameters were excluded from the study. This step is crucial as our method relies fundamentally on the geometric relationship between the planet and its star, and the star's thermal properties. Systems with high uncertainty flags or non-numeric values were also filtered out to ensure the statistical integrity of our sample. This curation process resulted in a final, high-quality dataset of several thousand exoplanets, which served as the basis for all subsequent classification and analysis.

\subsection{Stellar Classification and Sample Selection}
A central tenet of our hypothesis is that the habitability filter is dependent on the host star's spectral class. Using the curated dataset, we segregated the host stars into distinct bins based on their effective temperature ($T_*$), adhering to the standard Morgan-Keenan (MK) classification boundaries. Our analysis focused on the three stellar classes most commonly considered in the search for habitable worlds:
\begin{itemize}
    \item \textbf{G-type (Sun-like stars):} $5000 \, \text{K} \leq T_* < 6000 \, \text{K}$
    \item \textbf{K-type (Orange dwarfs):} $3500 \, \text{K} \leq T_* < 5000 \, \text{K}$
    \item \textbf{M-type (Red dwarfs):} $T_* < 3500 \, \text{K}$
\end{itemize}
This categorization allowed for an independent analysis of the planet populations orbiting each type of star. An initial investigation into planets orbiting hotter F-type stars ($T_* \geq 6000 \, \text{K}$) was conducted. However, consistent with discussions in the literature regarding the challenges to habitability around such stars (e.g., shorter main-sequence lifetimes and higher UV radiation flux \citep{Sato2017}), our method did not yield a significant correlation. Consequently, F-type systems, along with the even hotter and shorter-lived A, B, and O-type stars, were excluded from our primary analysis to focus on the stellar types most likely to host long-term habitable environments.

\subsection{Calculation of a Uniform Earth Similarity Index (ESI)}
To validate our geometric filter, a consistent, quantitative measure of habitability was required for comparison. For this, we utilized the Earth Similarity Index (ESI), a widely used metric that scores a planet's similarity to Earth on a scale from 0 to 1 \citep{SchulzeMakuch2011}. While the NASA Exoplanet Archive provides ESI values for some planets, they are not universally available across the entire catalog. To ensure uniformity, we implemented a custom algorithm to calculate the ESI for every planet in our curated dataset. The ESI is a function of multiple parameters, but for exoplanet studies, it is often simplified based on the available data. Our calculation is based on the two most robustly determined properties: planetary radius and the incident stellar flux, using the formula:
\begin{equation}
    ESI(S,R) = 1 - \sqrt{\frac{1}{2}\left[\left(\frac{S-S_{\oplus}}{S+S_{\oplus}}\right)^2 + \left(\frac{R-R_{\oplus}}{R+R_{\oplus}}\right)^2\right]}
\end{equation}
where $S$ and $R$ are the incident stellar flux and radius of the exoplanet, and $S_{\oplus}$ and $R_{\oplus}$ are Earth's values (1 solar flux and 1 Earth radius, respectively). The incident flux $S$ for each planet was calculated from the stellar luminosity and orbital distance. This approach, while not a complete measure of habitability, provides a standardized benchmark to effectively test the performance of our proposed filter.

\subsection{Development of the Tiered d/D\textsubscript{s} Ratio Filter}
The core of our methodology is the development of a novel first-order filter based on the dimensionless ratio of a planet's orbital semi-major axis to its host star's diameter ($d/D_s$). Our initial hypothesis, based on Earth's unique position in the solar system, was that a single, universal $d/D_s$ ratio might exist for all habitable planets. However, preliminary analysis quickly revealed that this was not the case; planets considered potentially habitable were found at systematically different ratios around different types of stars.

This led to the development of our tiered system. We identified a "gold-standard" benchmark planet for each of the G, K, and M spectral classes, selecting planets with the highest known ESI values that are considered archetypes for potentially habitable worlds.
\begin{itemize}
    \item For \textbf{G-type} stars, Earth is the unambiguous benchmark ($ESI = 1.0$), being the only known inhabited planet. Its $d/D_s$ ratio is approximately 107. We adopted an idealized benchmark ratio of \textbf{108} for this class.
    \item For \textbf{K-type} stars, we selected Kepler-442b ($ESI = 0.84$), one of the most robustly validated and Earth-like planets known to orbit a K-dwarf \citep{Torres2015}. Its native $d/D_s$ ratio is approximately 68. We observed this to be roughly half of the G-type standard, leading us to adopt an idealized ratio of \textbf{54}.
    \item For \textbf{M-type} stars, Teegarden's Star b ($ESI = 0.95$) was chosen as an exemplary benchmark, being a high-ESI Earth-mass planet orbiting a nearby M-dwarf \citep{Zechmeister2019}. Its native ratio is approximately 23. This value is roughly half of the K-type standard, leading to an idealized ratio of \textbf{27}.
\end{itemize}
This distinct empirical pattern—a halving of the ideal $d/D_s$ ratio with each step down in stellar temperature from G to K to M—forms the predictive basis of our filter. In the subsequent Results section, we test this tiered model by plotting $d$ versus $D_s$ for the planet population in each class and drawing lines of constant slope corresponding to these idealized ratios. We then analyze the planets that cluster along these "habitability sequences" to validate the filter's effectiveness.

\section{Results}
\label{sec:results}

The application of our tiered $d/D_s$ filter to the curated exoplanet dataset yielded distinct and compelling results for each of the G, K, and M stellar classes. The analysis demonstrates a strong correlation between our proposed habitability sequences and a high Earth Similarity Index (ESI), with the effectiveness of the filter increasing for cooler stars.

For the population of planets orbiting G-type stars, our filter, corresponding to a $d/D_s$ ratio of 108, successfully identified a distinct population of planets clustered near this line. This group is separate from the bulk of known exoplanets, which are typically "hot Jupiters" that orbit much closer to their host stars. Our analysis shows that among the planets identified by our filter, 50\% have an ESI greater than 0.5, indicating a significant success rate in identifying planets with more Earth-like characteristics, including candidates such as Kepler-452b (see Table ~\ref{tab:g_type_data}).

The correlation was even more pronounced for K-type stellar systems. Using the idealized $d/D_s$ ratio of 54, we found that 87.5\% of the planets clustered around this habitability line possess an ESI greater than 0.56. This high success rate suggests that our filter is particularly well-tuned for identifying promising candidates around K-type stars, such as Kepler-62f and Kepler-283c (see Table ~\ref{tab:k_type_data}).

The strongest validation of our tiered approach was observed for M-type stellar systems. With a habitability sequence defined by $d/D_s = 27$, we found that 100\% of the planets situated on or near this line in our sample have an ESI greater than 0.6. The filter effectively isolates all the well-known, high-ESI M-dwarf planets, including those in the TRAPPIST-1 system and Proxima Centauri b (see Table~\ref{tab:m_type_data}).

\begin{figure}[h!]
    \centering
    \includegraphics[width=0.5\textwidth]{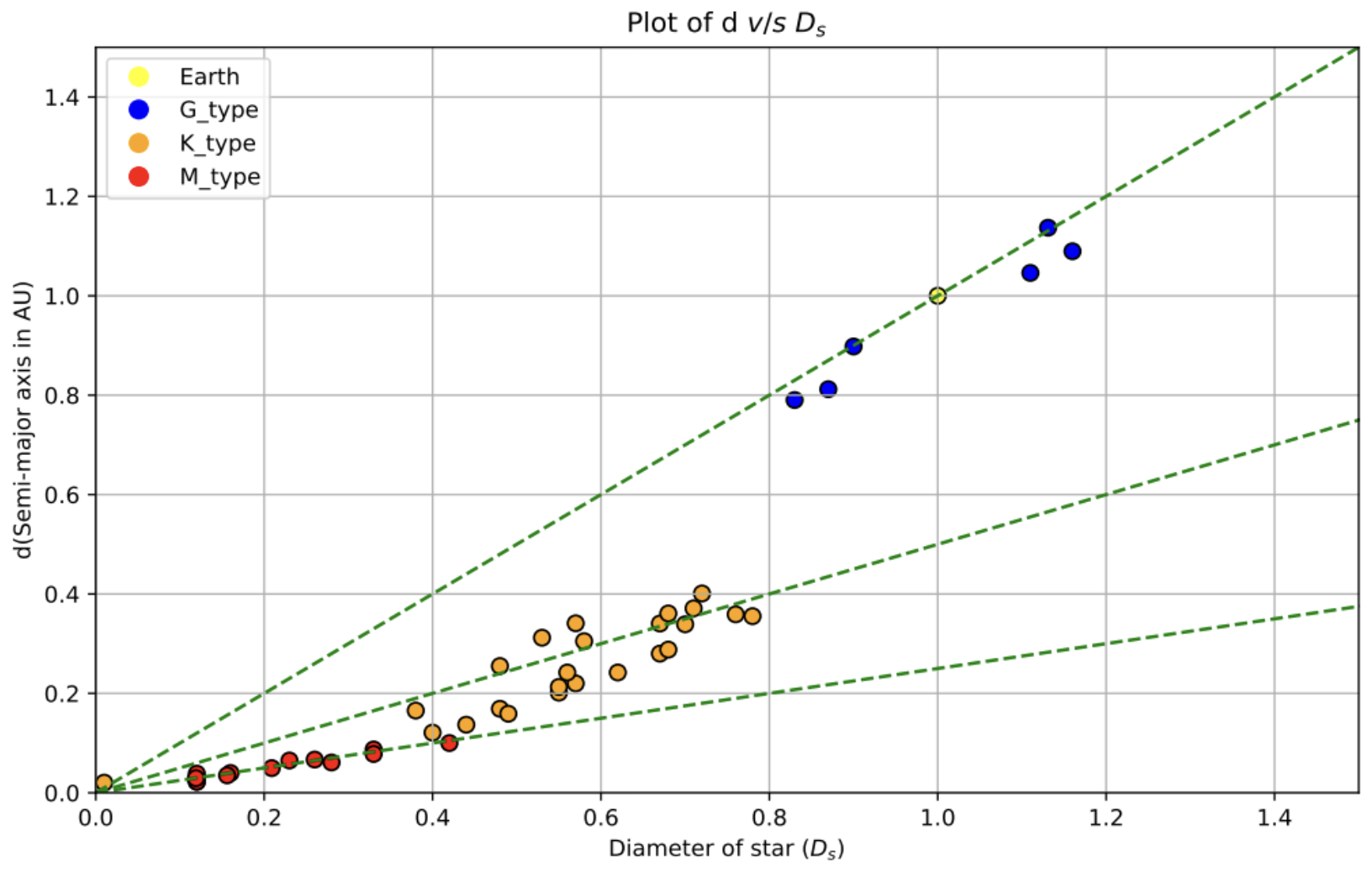}
    \caption{A composite plot showing the idealized habitability sequences for G-type (top), K-type (middle), and M-type (bottom) stars. This figure encapsulates the tiered nature of the proposed filter and serves as an analog to the H-R diagram for exoplanet habitability.}
    \label{fig:combined_plot}
\end{figure}

To visualize the complete tiered model, the habitability sequences for all three stellar classes are presented on a single composite plot in Figure \ref{fig:combined_plot}.

\section{Discussion}
\label{sec:discussion}

\subsection{Interpretation of the Tiered Habitability Sequences}
The results presented in this work strongly support our central hypothesis: that the geometric ratio of a planet's orbital distance to its star's diameter ($d/D_s$) is a potent first-order indicator of potential habitability, provided it is tiered according to the star's spectral class. The clear delineation of planets with high ESI values along the proposed habitability sequences of $d/D_s = 108, 54,$ and $27$ for G, K, and M-type stars, respectively, is not a mere coincidence. It is a direct geometric consequence of the fundamental principles of stellar physics and the definition of the habitable zone.

As established in our theoretical framework, for a planet to maintain a surface temperature conducive to liquid water, it must receive a specific amount of energy flux from its host star. M-dwarfs, being significantly cooler and less luminous than G-type stars like our Sun, have habitable zones that are much more compact. A planet must orbit substantially closer to an M-dwarf to receive the same amount of energy as Earth receives from the Sun. Our $d/D_s$ ratio elegantly encapsulates this relationship. The observed factor-of-two decrease in the ratio from G to K and from K to M is an empirical reflection of how the habitable zone's location scales with the star's fundamental properties. It confirms that the system's overall geometry is a powerful proxy for the more complex physics of stellar energy output and orbital mechanics.

\subsection{An Analog to the Hertzsprung-Russell Diagram for Exoplanets}
One of the most powerful conceptual outcomes of this work is the creation of a diagram (Figure \ref{fig:combined_plot}) that serves as a practical analog to the Hertzsprung-Russell (H-R) diagram for exoplanet habitability. The H-R diagram revolutionized stellar astronomy by revealing that stars are not randomly distributed by luminosity and temperature, but fall along distinct sequences. Similarly, our plot shows that potentially habitable planets do not occupy a random parameter space. Instead, they appear to fall along well-defined "Habitability Main Sequences" based on their host star.

This provides astronomers with a powerful new heuristic. By simply calculating the $d/D_s$ ratio and plotting a newly discovered planet on this diagram, one can gain an immediate, intuitive sense of its potential habitability. Planets falling far from these lines are unlikely to be primary targets in the search for life, while those falling on or near a sequence can be immediately prioritized for more intensive follow-up observations. This graphical tool simplifies the complex task of initial candidate vetting from large-scale surveys like TESS \citep{Ricker2015}.

\subsection{Limitations and Avenues for Future Research}
It is crucial to acknowledge that the proposed method is a \textit{first-order filter}. Habitability is a multifaceted concept that depends on a host of factors not considered in our geometric approach. These include, but are not limited to:
\begin{itemize}
    \item \textbf{Planetary Mass and Composition:} Our filter is agnostic to whether a planet is rocky or gaseous. A giant planet could fall on a habitability sequence, but would not be considered habitable (though its moons might be). Future work could incorporate planetary density as a second-layer filter.
    \item \textbf{Atmospheric Properties:} The presence, composition, and density of an atmosphere are critical for maintaining surface water and regulating temperature. Our filter serves to identify candidates where such an atmosphere \textit{could} exist. The crucial task of atmospheric characterization remains the domain of advanced spectroscopic instruments like the James Webb Space Telescope (JWST) \citep{Gardner2006}.
    \item \textbf{Tidal Locking:} Planets orbiting close to M-dwarfs, as many on our M-type sequence do, are likely to be tidally locked. While this was once considered a barrier to habitability, modern climate models suggest that atmospheres could still support liquid water under these conditions \citep{Kopparapu2013}. Our filter identifies these candidates, but does not resolve the complexities of their climate states.
\end{itemize}
The primary utility of our filter is therefore not to provide a definitive assessment of habitability, but to dramatically narrow the field of candidates. It is a tool for target selection, enabling the efficient allocation of precious observational resources to the planets with the highest probability of being Earth-like. 

\section{Conclusion}
\label{sec:conclusion}

In response to the ever-growing catalog of exoplanets, we have developed and validated a novel, efficient, and physically motivated first-order filter to identify potentially habitable worlds. The vast datasets produced by missions like Kepler and TESS necessitate heuristic tools to triage and prioritize candidates for further study. Our work provides such a tool, grounded in a simple geometric ratio.

The core contribution of this paper is the demonstration that the dimensionless ratio of a planet's semi-major axis to its host star's diameter ($d/D_s$) is a powerful predictor of habitability, but only when applied in a tiered system dependent on the star's spectral class. We have established three idealized "habitability main sequences" defined by specific ratios:
\begin{itemize}
    \item \textbf{G-type stars:} $d/D_s \approx 108$
    \item \textbf{K-type stars:} $d/D_s \approx 54$
    \item \textbf{M-type stars:} $d/D_s \approx 27$
\end{itemize}
Our analysis shows that planets from the NASA Exoplanet Archive that conform to these geometric constraints exhibit a remarkably high correlation with a large Earth Similarity Index (ESI), with the predictive power of the filter increasing for cooler stars.

We propose that the composite plot of these sequences can serve the exoplanet community as an intuitive and practical analog to the H-R diagram, allowing for the rapid classification of new discoveries. While this filter does not replace the need for detailed follow-up analysis of planetary atmospheres and composition, it provides a vital preliminary step. By efficiently shortlisting the most promising candidates from thousands of possibilities, this method can help focus our observational resources and accelerate the ongoing search for a true Earth 2.0.

\section{Acknowledgement}\label{sec: Acknowledgement}

The authors wish to express their sincere gratitude to the Department of Physics at Fergusson College (Autonomous), Pune. The conducive research atmosphere and institutional support provided were essential for the completion of this project. 

We are particularly indebted to Dr. Priyanka Chaturvedi of the Tata Institute of Fundamental Research (TIFR); her valuable insights and constructive discussions during the nascent stages of this work were instrumental in shaping its initial direction and refining our core hypothesis. 

This research has made extensive use of the NASA Exoplanet Archive, which is operated by the California Institute of Technology, under contract with the National Aeronautics and Space Administration under the Exoplanet Exploration Program. The open accessibility of this invaluable dataset is fundamental to progress in the field.

\appendix

\section{Tabulated Data for High-ESI Exoplanet Candidates}
\label{sec:appendix_data}
\
This appendix provides the detailed data for the full sample of exoplanets analysed in this study and visualised in Figure \ref{fig:combined_plot}. The tables below categorise these planets by the spectral type of their host star (G, K, and M). For each exoplanet, we provide the semi-major axis of its orbit ($d$) in AU, the diameter of its host star ($D_s$) in solar unite, and its calculated Earth Similarity Index (ESI). These parameters form the basis of our analysis. All data were sourced from the NASA Exoplanet Archive \citep{Akeson2013}.

\begin{table}[h!]
\centering
\small
\setlength{\tabcolsep}{3pt}
\caption{G-type Exoplanet Candidates \citep{Akeson2013}}
\label{tab:g_type_data}
\begin{tabular}{l l c c c}
\hline
\textbf{Planet Name} & \textbf{Host Star} & \textbf{d (AU)} & \textbf{$D_s$ (Solar)} & \textbf{ESI} \\
\hline
Earth                & Sun                & 1.000           & 1.000          & 1.000         \\
Kepler-34 b          & Kepler-34          & 1.090           & 1.160          & 0.436         \\
Kepler-453 b         & Kepler-453         & 0.790           & 0.830          & 0.488         \\
Kepler-553 c         & Kepler-553         & 0.898           & 0.900          & 0.386         \\
Kepler-22 b          & Kepler-22          & 0.812           & 0.870          & 0.749         \\
Kepler-452 b         & Kepler-452         & 1.046           & 1.110          & 0.828         \\
KOI-4878.01          & KOI-4878           & 1.137           & 1.331          & 0.983         \\
\hline
\end{tabular}
\end{table}

\begin{table}[h!]
\centering
\small
\setlength{\tabcolsep}{3pt}
\caption{K-type Exoplanet Candidates \citep{Akeson2013}}
\label{tab:k_type_data}
\begin{tabular}{l l c c c}
\hline
\textbf{Planet Name} & \textbf{Host Star} & \textbf{d (AU)} & \textbf{$D_s$ (Solar)} & \textbf{ESI} \\
\hline
Kepler-186 f         & Kepler-186         & 0.432           & 0.520          & 0.577         \\
TOI-964 c            & TOI-964            & 0.332           & 0.530          & 0.854         \\
WD 1856+534 b        & WD 1856+534        & 0.020           & 0.010          & 0.183         \\
Kepler-442 b         & Kepler-442         & 0.409           & 0.600          & 0.843         \\
Kepler-62 f          & Kepler-62          & 0.427           & 0.640          & 0.826         \\
Kepler-1653 b        & Kepler-1653        & 0.471           & 0.690          & 0.788         \\
Kepler-1652 b        & Kepler-1652        & 0.165           & 0.380          & 0.822         \\
Kepler-309 c         & Kepler-309         & 0.401           & 0.720          & 0.879         \\
Kepler-283 c         & Kepler-283         & 0.341           & 0.570          & 0.791         \\
Kepler-298 d         & Kepler-298         & 0.305           & 0.580          & 0.685         \\
Kepler-296 f         & Kepler-296         & 0.255           & 0.480          & 0.737         \\
Kepler-443 b         & Kepler-443         & 0.499           & 0.710          & 0.715         \\
\hline
\end{tabular}
\end{table}

\begin{table}[h!]
\centering
\small
\setlength{\tabcolsep}{3pt}
\caption{M-type Exoplanet Candidates \citep{Akeson2013}}
\label{tab:m_type_data}
\begin{tabular}{l l c c c}
\hline
\textbf{Planet Name} & \textbf{Host Star} & \textbf{d (AU)} & \textbf{$D_s$ (Solar)} & \textbf{ESI} \\
\hline
TOI-4336 A b       & TOI-4336 A       & 0.087           & 0.330          & 0.709         \\
Gliese 12 b        & Gliese 12        & 0.067           & 0.260          & 0.894         \\
TRAPPIST-1 e       & TRAPPIST-1       & 0.029           & 0.120          & 0.845         \\
TRAPPIST-1 f       & TRAPPIST-1       & 0.038           & 0.120          & 0.677         \\
LP 890-9 c         & LP 890-9         & 0.040           & 0.160          & 0.888         \\
TOI-1452 b         & TOI-1452         & 0.061           & 0.280          & 0.715         \\
TRAPPIST-1 d       & TRAPPIST-1       & 0.023           & 0.120          & 0.908         \\
K2-72 c            & K2-72            & 0.078           & 0.330          & 0.755         \\
TOI-198 b          & TOI-198          & 0.100           & 0.420          & 0.720         \\
GJ 1061 c          & GJ 1061          & 0.035           & 0.156          & 0.709         \\
Ross 128 b         & Ross 128         & 0.050           & 0.209          & 0.746         \\
Teegarden's Star b & Teegarden's Star & 0.026           & 0.120          & 0.950         \\
Kepler-1649 c      & Kepler-1649      & 0.065           & 0.230          & 0.860         \\
Trappist-1 d       & Trappist-1       & 0.022           & 0.120          & 0.603         \\
TRAPPIST-1 e       & TRAPPIST-1 e     & 0.029           & 0.119          & 0.769         \\
\hline
\end{tabular}
\end{table}

\clearpage

\bibliography{References}{}

@article{Mayor1995,
  author  = {{Mayor}, M. and {Queloz}, D.},
  title   = "{A Jupiter-Mass Companion to a Solar-Type Star}",
  journal = {Nature},
  year    = {1995},
  volume  = {378},
  number  = {6555},
  pages   = {355-359},
  doi     = {10.1038/378355a0}
}

@article{Borucki2010,
  author  = {{Borucki}, W.~J. and {Koch}, D. and {Basri}, G. and {et al.}},
  title   = "{Kepler Planet-Detection Mission: Introduction and First Results}",
  journal = {Science},
  year    = {2010},
  volume  = {327},
  number  = {5968},
  pages   = {977-980},
  doi     = {10.1126/science.1185402}
}

@article{Ricker2015,
  author  = {{Ricker}, G.~R. and {Winn}, J.~N. and {Vanderspek}, R. and {et al.}},
  title   = "{Transiting Exoplanet Survey Satellite (TESS)}",
  journal = {Journal of Astronomical Telescopes, Instruments, and Systems},
  year    = {2015},
  volume  = {1},
  eid     = {014003},
  pages   = {014003},
  doi     = {10.1117/1.JATIS.1.1.014003}
}

@article{Kasting1993,
  author  = {{Kasting}, J.~F. and {Whitmire}, D.~P. and {Reynolds}, R.~T.},
  title   = "{Habitable Zones around Main Sequence Stars}",
  journal = {Icarus},
  year    = {1993},
  volume  = {101},
  number  = {1},
  pages   = {108-128},
  doi     = {10.1006/icar.1993.1010}
}

@article{SchulzeMakuch2011,
  author  = {{Schulze-Makuch}, D. and {Méndez}, A. and {Fairén}, A.~G. and {et al.}},
  title   = "{A Two-Tiered Approach to Assess the Habitability of Exoplanets}",
  journal = {Astrobiology},
  year    = {2011},
  volume  = {11},
  number  = {10},
  pages   = {1041-1052},
  doi     = {10.1089/ast.2010.0592}
}

@book{Pierrehumbert2010,
  author    = {{Pierrehumbert}, R.~T.},
  title     = "{Principles of Planetary Climate}",
  publisher = {Cambridge University Press},
  year      = {2010},
  doi       = {10.1017/CBO9780511780783}
}

@book{Kippenhahn2012,
  author    = {{Kippenhahn}, R. and {Weigert}, A. and {Weiss}, A.},
  title     = "{Stellar Structure and Evolution}",
  series    = {Astronomy and Astrophysics Library},
  publisher = {Springer-Verlag Berlin Heidelberg},
  year      = {2012},
  doi       = {10.1007/978-3-642-30304-3}
}

@article{Akeson2013,
  author  = {{Akeson}, R.~L. and {Chen}, X. and {Ciardi}, D. and {et al.}},
  title   = "{The NASA Exoplanet Archive: Data and Tools for Exoplanet Research}",
  journal = {Publications of the Astronomical Society of the Pacific},
  year    = {2013},
  volume  = {125},
  number  = {930},
  pages   = {989},
  doi     = {10.1086/672273}
}

@article{Torres2015,
  author  = {{Torres}, G. and {Kipping}, D.~M. and {Fressin}, F. and {et al.}},
  title   = "{Validation of 12 Small Kepler Transiting Planets in the Habitable Zone}",
  journal = {The Astrophysical Journal},
  year    = {2015},
  volume  = {800},
  number  = {2},
  pages   = {99},
  doi     = {10.1088/0004-637X/800/2/99}
}

@article{Zechmeister2019,
  author  = {{Zechmeister}, M. and {Dreizler}, S. and {Ribas}, I. and {et al.}},
  title   = "{The CARMENES search for exoplanets around M dwarfs. Two temperate Earth-mass planet candidates around Teegarden's Star}",
  journal = {Astronomy \& Astrophysics},
  year    = {2019},
  volume  = {627},
  eid     = {A49},
  pages   = {A49},
  doi     = {10.1051/0004-6361/201935460}
}

@article{Sato2017,
  author = {{Sato}, S. and {Cuntz}, M. and {Menou}, K. and {Jack}, D.},
  title = "{Habitable Zones around F- and G-type Stars against the Influence of UV Radiation}",
  journal = {The Astrophysical Journal},
  year = {2017},
  volume = {848},
  number = {2},
  pages = {124},
  doi = {10.3847/1538-4357/aa8a51}
}

@article{Gardner2006,
  author  = {{Gardner}, J.~P. and {Mather}, J.~C. and {Clampin}, M. and {et al.}},
  title   = "{The James Webb Space Telescope}",
  journal = {Space Science Reviews},
  year    = {2006},
  volume  = {123},
  number  = {4},
  pages   = {485-606},
  doi     = {10.1007/s11214-006-8315-7}
}

@article{Kopparapu2013,
  author  = {{Kopparapu}, R.~K. and {Ramirez}, R. and {Kasting}, J.~F. and {et al.}},
  title   = "{Habitable Zones around Main-sequence Stars: New Estimates}",
  journal = {The Astrophysical Journal},
  year    = {2013},
  volume  = {765},
  number  = {2},
  pages   = {131},
  doi     = {10.1088/0004-637X/765/2/131}
}
\bibliographystyle{aasjournal}



\end{document}